\documentclass[10pt]{article}
\usepackage{geometry}                % See geometry.pdf to learn the layout options. There are lots.
\usepackage{graphicx}
\usepackage{amssymb}
\usepackage{epstopdf}
\usepackage{amsmath}

\usepackage{subfigure}

\title{Realistic Boundary Conditions for Ad Hoc Network Node Mobility Models and a New Approach to the Random Waypoint Model}
\author{Michael Hohensee}
\begin{document}
\maketitle

%\section{}
%\subsection{}
\begin{abstract}
In this paper, we examine the cause of the border effect observed in many mobility models used to construct simulations of ad hoc networking protocol performance.  We specify conditions under which a node mobility model must produce spatial mobile node  distribution functions that obey the diffusion equation.  In particular demonstrate that these conditions are satisfied by the random direction (RD) model.  We show that it is possible to construct mobility models that attain uniform steady-state distributions without resorting to reflection or ``wrapping'' of nodes at the border of a test region.  Finally, we show that the random waypoint (RWP) model may be reproduced by the application of a ``volume rule'' to an RD model.  This volume rule violates the assumptions that lead to the diffusion equation.  We suggest a generalization of the RWP model that can provide more uniform mobile node distributions.
\end{abstract}

\section{Introduction}
At present, most wireless devices rely upon fixed infrastructure to connect to one another and the outside world.  Fixed installations are necessary due to the need for power to operate them, access to a larger wired network, and for centralized processing to coordinate relatively unsophisticated devices.  As the number and sophistication of portable wireless network devices increases, it becomes useful to consider replacing this centralized network architecture with one that is generated spontaneously among local network nodes, or an ad hoc network.  Ad hoc networking promises to liberate wireless devices from dependence on fixed infrastructure by allowing devices to communicate to one another via a dynamically generated network, where peers pass data to and through one another rather than to a fixed central node.  In order to achieve this, network protocols (NPs) must be developed that allow independent nodes to agree upon a network topology, and that can then keep pace with changes to the network as nodes move in and out of range of one another.  

In recent years, there has been much interest in designing models for simulating the operation of arbitrary ad hoc networking protocols for networks with mobile nodes.  Such simulations are typically defined by:
\begin{enumerate}
\item The network protocol to be tested, or a specified property of the mobile node ensemble to be studied, such as internode distances, the existence of spanning trees subject to relevant constraints, etc.
\item A fixed test volume of $d$-dimensional space (often with $d=2$) in which mobile nodes move about.
\item A mobility model that specifies the behavior and motion of mobile nodes as they move about the volume, which may include rules describing the conditions under which mobile nodes may exit the test volume (and hence the simulation) or be introduced into it.
\end{enumerate}

In most cases, network protocols are not designed for a particular location, where the locations of obstacles and barriers are known in advance.  Instead, we would like to construct protocols that perform well in general circumstances, and need to quantify the performance of these protocols knowing only the mobility model that mobile nodes obey and the average node density.  The relative position of the nodes has a significant impact on a protocol's performance, yet we do not usually know the shape of the volume in which the mobile nodes will be confined.  The way we proceed is to define arbitrary test volumes, which are meant to be representative of portions of the actual volume in which mobile nodes are deployed.  We then expect the nodes to move without particular regard for the specific shape of the test volume.  Thus, a key feature of a mobile node simulator should be that the distribution of nodes within the test volume be independent of that volume's shape.

It has been noted that for most mobility models in common use, \cite{Bettstetter:2001,Santi:2005}, the steady-state mobile node spatial distribution function is nonuniform over the test volume, and a function of the test volume geometry \cite{Bettstetter:2002}.  These density gradients can arise from rules restricting nodes from the border, as in the random waypoint (RWP) model, or from the manner in which nodes that exit the test volume are replaced.  Nonuniformities in the steady-state mobile node distribution can have a significant impact on simulation-based estimates of the likelihood that a given NP fails due to a lack or overabundance of node connectivity.  It is common for estimates of conditions under which a given network protocol can function to vary significantly for different mobility models with equal numbers of mobile nodes.  These discrepancies are partially due to model-specific density gradients over the test volume, which artificially increase the mobile node density in some areas while decreasing it elsewhere.  It is important that we be able to distinguish real effects that are due to the local motion of mobile nodes from artifacts or ``border effects'' generated by the geometry of our simulation.

A significant amount of effort has been devoted to understanding the steady-state distributions associated with various mobility models \cite{Camp:2002}, and the conditions under which the distributions approximate uniform distribution functions \cite{Blough:2004}.  To date, work has focused on the results of direct numerical simulations of mobile node motion for specific models.  In section \ref{sec:diffusion}, we demonstrate that the evolution of the mobile node distribution function often obeys the diffusion equation, and provide a general analytic method for specifying and studying the distribution function.

The existence of a mobile node density gradient can give rise to threshold effects \cite{Santi:2005a}, which occur when the number of mobile nodes increases to the point that it becomes likely that at least one of them will visit relatively depleted regions.  These gradients have been demonstrated to vanish for models that cause mobile nodes to be reflected, turned back, or wrapped to the opposite side of the test region at the boundary \cite{Bettstetter:2001}.  Unfortunately, these border rules may not be realistic for every class of mobile node, on either behavioral or topological grounds.  These border rules generate uniform mobile node distributions because they ensure that, for any infinitesimal surface element on the boundary, the rate at which mobile nodes exit the test volume through the surface element is equal to the rate at which they are introduced through that element.  This is the same as saying that the mobile node flux through the border satisfies the principle of detailed balance \cite{Pathria:1996a}, or that the system of mobile nodes in the test volume is in equilibrium with the surrounding environment.  Section \ref{sec:distsec} of this paper introduces a more realistic way to satisfy the principle of detailed balance on the border for arbitrary mobility models, using a random direction (RD) model as an example.  mobile nodes are introduced on the border at points sampled from a distribution that matches the distribution of points through which mobile nodes exit the test region.

In section \ref{sec:rwp}, we consider the case of the random waypoint (RWP) mobility model.  We offer a simple explanation of the observation in \cite{Blough:2004} that the uniformity of the steady-state mobile node distribution function is independent of the maximum mobile node velocity, and why it varies with the probability that mobile nodes are stationary, and the mean delay before mobile nodes move.  We show that the RWP model can never yield uniform distributions of mobile nodes.  Recasting the RWP model as a special case of an RD model leads us to suggest new ways to improve the spatial uniformity of the steady-state mobile node distribution.

\section{The Diffusion Equation\label{sec:diffusion}}
A famous mobility model is the random walk (RW) model.  Also known as Brownian motion, it is known to produce a particle (or mobile node) distribution $f(x,t)$ that obeys the diffusion equation,
\begin{equation}
\frac{\partial f(x,t)}{\partial t} = D\nabla^{2}f(x,t)\label{eq:diffusionEq}.
\end{equation}
The diffusion equation is applicable to a wide variety of physical systems, requiring only that the following four conditions be satisfied:
\begin{enumerate}
\item{The probability density function for the position of an arbitrary node must be normalized for all time $t$, that is
\begin{equation}
\int_{V}f(x,t) = 1
\end{equation}\label{cond:one}}
\item The probability density function $W(\boldsymbol{\xi})$ for a node to be displaced by $\boldsymbol{\xi}$ in a short time $\delta t$ must be sharply peaked about $\boldsymbol{\xi}=\mathbf{0}$, and independent of the position $\mathbf{x}$ about which the node is to be displaced.\label{cond:two}  By ``sharply peaked'', we mean that the variance in the distribution of node displacements must increase linearly with time.  If this variance varies as $t^{\gamma}$, where $\gamma<1$ or $\gamma>1$, sub-diffusive or super-diffusive behavior will result.
\item $W(\boldsymbol{\xi})=W(-\boldsymbol{\xi})$, indicating that the mobile node is not biased towards motion in any particular direction, and so the mean displacement is zero.\label{cond:three}
\item $W(\boldsymbol{\xi})$ must be factorizable into $W_{1}(\xi_{1})\dots W_{d}(\xi_{d})$ for a $d$-dimensional volume.  

In other words, the probability that after time $\delta t$ the mobile node is displaced by $\xi_{i}$ along one spatial degree of freedom is independent of the probability of its being displaced 
its displacement along the other spatial degrees of freedom.\label{cond:four}
\end{enumerate}
This derivation has been adapted from \cite{Pathria:1996}.  If the above conditions are satisfied, we may write 
\begin{equation}
\frac{\partial f(x,t)}{\partial t} = \int_{V}d\boldsymbol{\xi}\left[f(\mathbf{x}+\boldsymbol{\xi},t)W(\boldsymbol{-\xi})-f(\mathbf{x},t)W(\boldsymbol{\xi})\right]\label{eq:master}.
\end{equation}
The first term on the right-hand side of (\ref{eq:master}) describes the rate at which mobile nodes move to the small volume about $\mathbf{x}$ from other positions $\mathbf{x}+\boldsymbol{\xi}$, while the second term accounts for the rate at which mobile nodes leave the small volume about $\mathbf{x}$.  As $W(\boldsymbol{\xi})$ is sharply peaked about $0$, the integral in (\ref{eq:master}), may be approximated by a Taylor expansion about $\boldsymbol{\xi}=\mathbf{0}$.
\begin{equation}
\frac{\partial f(x,t)}{\partial t} = \frac{1}{2}\nabla^{2}\left[\int_{V}\xi^{2}W(\boldsymbol{\xi})d\boldsymbol{\xi}f(x,t)\right],
\end{equation}
where we have taken advantage of conditions \ref{cond:three} and \ref{cond:four} to eliminate first order terms in the Taylor expansion as well as terms proportional to $\frac{\partial^{2}}{\partial \xi_{i}\partial \xi_{j}}$ for $i\neq j$.  With condition \ref{cond:two}, we obtain the diffusion equation (\ref{eq:diffusionEq}), where $D$ can, by applying the central limit theorem, be expressed as:
\begin{equation}
D = \frac{\langle \xi^{2}\rangle}{\tau},
\end{equation}
and $\langle \xi^{2}\rangle$ is the variance in the average displacement after a typical movement period $\tau$.  

\section{Uniformity in the Random Direction Model\label{sec:distsec}}

In this section, we define the random direction (RD) model, and show that the mobile node distribution it generates satisfies the diffusion equation.  We then review three common techniques used to compensate for the loss of mobile nodes as they diffuse beyond the boundary of the test region, two of which generate uniform steady-state spatial mobile node distributions.  We introduce a fourth method that has heretofore been overlooked, and that we think is more realistic.  By construction, our method is generalizable to arbitrary mobile node mobility models that allow mobile nodes to freely traverse the border of a finite test region.

An RD model is defined by a set of parameters $v_{min}$, $v_{max}$, $\tau_{s}$, and $\tau_{m}$, which are, respectively, the minimum node velocity, the maximum node velocity, an average pause time, and the average time that a node will move in a step.
The motion of the mobile nodes is determined according to the following model:
\begin{quote}
RD Mobility Model

At the beginning of a step, each node (a) chooses a set $\{\theta,t_{s},t_{m},v\}$, where $\theta$ and $v$ are drawn uniformly at random from the respective intervals $[0,2\pi)$, and $[v_{min},v_{max}]$, and $\{t_{s},t_{m}\}$ are drawn from exponential distributions with means $\tau_{s}$ and $\tau_{m}$, respectively.  (b) The node delays for $t_{s}$ seconds, and then moves in the direction indicated by $\theta$ with speed $v$ for a time $t_{m}$, after which the step is complete, and the node repeats (a).
\end{quote}

The RD model satisfies the four diffusion equation conditions of section \ref{sec:diffusion}.  The number of mobile nodes is conserved in the test region, satisfying condition \ref{cond:one}.  Nodes move continuously from place to place, and so for sufficiently short times, the probability that mobile nodes will move a distance $\xi$ from their initial positions is sharply peaked about zero, satisfying condition \ref{cond:two}.  The direction in which the mobile nodes move is selected from a uniform distribution, which implies that $W(\xi)=W(-\xi)$, satisfying condition \ref{cond:three}, and that the probability of displacement along one degree of freedom does not correlate with the probability of displacement along other degrees of freedom, satisfying condition \ref{cond:four}.  Indeed, most mobility models that allow mobile nodes to choose direction uniformly at random (i.e., uncorrelated with mobile node position), and requires mobile nodes to move continuously from place to place will satisfy these conditions.

Given that the RD model obeys the diffusion equation, mobile nodes moving according to the RD model will inevitably exit any finite test region after a sufficient time.  Without some means of replenishing the number of mobile nodes in the test region, the only equilibrium distribution we might hope to produce is a distribution of zero mobile nodes.  A border rule \cite{Bettstetter:2001}, or boundary condition, is required to define what happens when a mobile node's trajectories intersect the boundary of the test region.  We shall consider commonly considered border rules where exiting mobile nodes are:  
\begin{enumerate}
\item reintroduced at a point selected uniformly at random within the test region (uniform replacement);
\item forced to choose a new direction that takes them back into the test region (inelastic reflection);
\item wrapped to the opposite side of the test region (simulation on a torus);
\end{enumerate}
and we shall introduce a new rule that calls for exiting mobile nodes to be
\begin{enumerate}
\item[4.] re-introduced at a point on the border drawn at random from a distribution that matches the distribution of points through which mobile nodes exit (random sampling).
\end{enumerate}

The first method allows mobile nodes to leave, and then recreates them at points distributed uniformly at random within the test region.  Because the RD model yields distributions that obey the diffusion equation, this method does not produce a uniform distribution of mobile nodes within the box.  The spatial distribution of mobile nodes which results has the same form as the distribution of temperatures of a uniformly heated box surrounded by a zero temperature reservoir.  The only difference is a normalization factor.  Because the node density distribution function (or temperature) is continuous, the chosen boundary conditions require it to drop to zero on the border.  In principle one could replace mobile nodes according to a nonuniform distribution, but in each case, the steady-state mobile node distribution function will mimic the temperature distribution on a similarly heated box surrounded by a zero-temperature reservoir.

The second method introduces a mobile node at the same point at which the exiting mobile node left, and draw sets of $\{\theta,t_{s},t_{m},v\}$ until one describes a non-escaping trajectory.  Because the motion of the mobile nodes is isotropic and not correlated with position within the box, the distribution of mobile node trajectories that leave the border at a given point is the same as the distribution of trajectories that land on that point.  We may think of this method as ``adding'' mobile nodes at the precise point in time and space that one ``leaves'' by intersecting the border.  As nodes are only added at points where nodes have just left, the rate at which nodes enter through a given point equals the rate at which they leave, resulting in a uniform distribution.

The third method introduces identical mobile nodes on the opposite side of the box, which would move according to the same set $\{\theta,t_{s},t_{m},v\}$, less whatever $t_{m}$ has already passed. This technique, which is applicable to boxes symmetric under reflection about their borders, produces uniform mobile node distributions because the distribution of mobile nodes entering the box is, due to symmetry, equal to the distribution of mobile nodes leaving the box.  The distribution that results is equivalent to a solution of the heat equation on a torus with some initial uniform temperature.  Because there is no external reservoir, this distribution remains uniform.

The final replacement method we will consider here, and focus on in this section, is to introduce mobile nodes at points along the border at random.  New nodes are assigned a pair $\{\mathbf{x},\theta\}$ drawn from a distribution with p.d.f. $g(\mathbf{x},\theta)$, where $\mathbf{x}$ is the point along the border the node is introduced, and $\theta$ is the direction it moves in with respect to the border.  This p.d.f. is chosen to be identical to the probability density for a mobile node exiting the box to leave through a point $\mathbf{x}$ at an angle $\theta$ with respect to the boundary.  This ensures a constant number of mobile nodes, and balances the mobile node flux entering the box with the flux leaving the box.

Because the mobile nodes are free to move across the border of the test area without being reflected or wrapped to the far side, this model allows for more realistic simulation of node mobility.  For some shapes and mobility models, the probability density $g(\mathbf{x},\theta)$ can be found analytically, but is in general nontrivial.  We derive the marginal p.d.f. $g(\mathbf{x})$ in appendix \ref{sec:analyticDist}.  For arbitrary test areas, this distribution may also be determined experimentally, by placing mobile nodes uniformly at random within the test area, running the mobility model through a single timestep, and noting the distribution of points on and the angles relative to the boundary at which the nodes leave.

\begin{figure}[ht]
\begin{center}
\mbox{
\subfigure[$v_{min}=0.001$m/s, $v_{max}=0.01$m/s]{\includegraphics[width=2.5in]{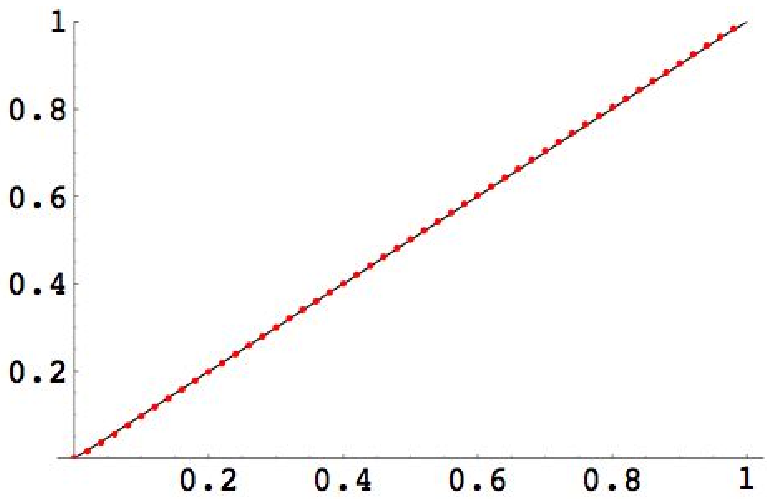}}\quad
\subfigure[$v_{min}=0.1$m/s, $v_{max}=1$m/s]{\includegraphics[width=2.5in]{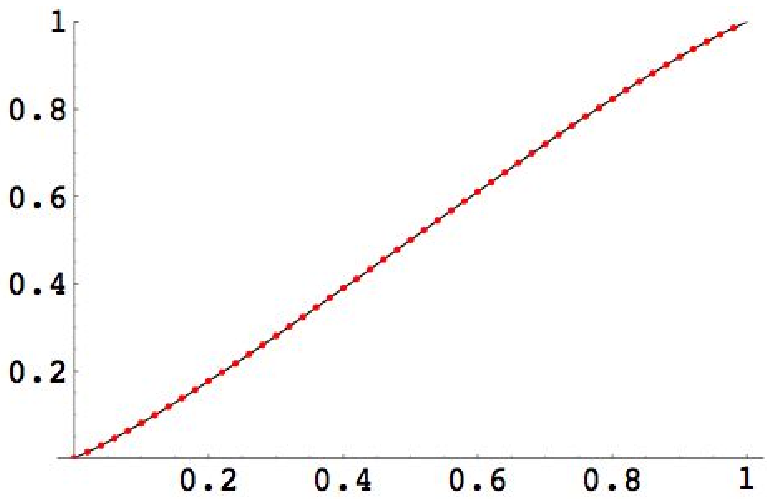}}
}
\end{center}
\caption{Simulated (points) vs calculated (line) cumulative distribution function $P(X\leq x)$ (vertical axes) for nodes to exit a distance $\leq x$ (horizontal axes) along a side of the $1\times1$ m$^{2}$ square.  Left: $\tau_{move}=3$s, $v_{min}=.001$m/s, $v_{max}=.01$m/s.  Right: $\tau_{move}=3$s, $v_{min}=.1$m/s, $v_{max}=1$m/s.}
\label{fig:RDDists}
\end{figure}

To test our border rule, we performed four simulations of $40,000$ mobile nodes, initially distributed uniformly on a unit $1$m$\times1$m square, for $9,900$ seconds of simulated time, choosing the range of velocities $v_{min}=0.001$m/s, $v_{max}=0.01$m/s, with an average pause time of $\tau_{pause}=1$s, and an average movement period of $\tau_{move}=3$s.  To determine the proper distribution for our new border rule, we simulate $100$ million mobile nodes taking a single step from points within the square distributed uniformly at random.  Because the square is symmetric, we need only determine the distribution for nodes to be introduced along one side.  Figure \ref{fig:RDDists} shows a typical marginal distribution function $P(X\leq x)$ that results.

\begin{figure}[ht] %  figure placement: here, top, bottom, or page
\begin{center}
\mbox{
	\subfigure[Uniform Replacement]{\includegraphics[width=1.5in]{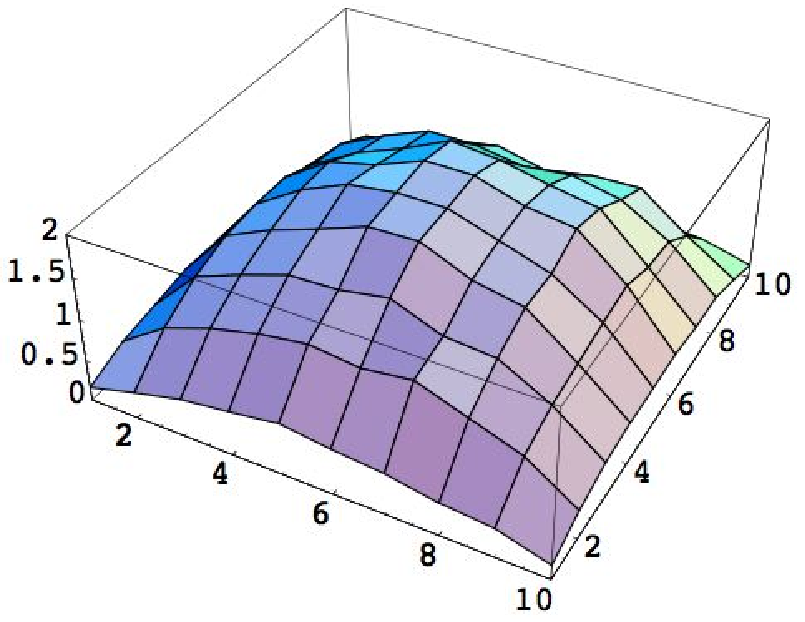}} \quad
	\subfigure[Reflection]{\includegraphics[width=1.5in]{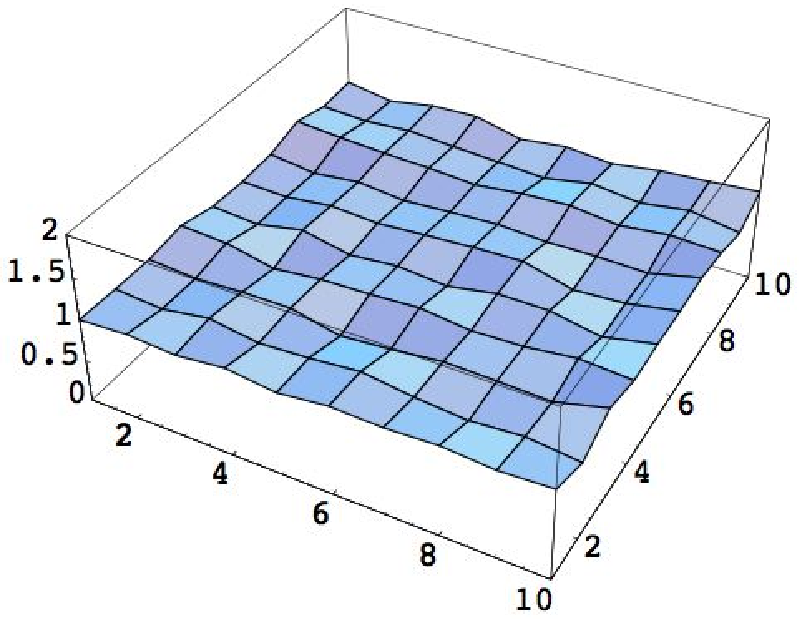}} \quad
	\subfigure[Wrapping]{\includegraphics[width=1.5in]{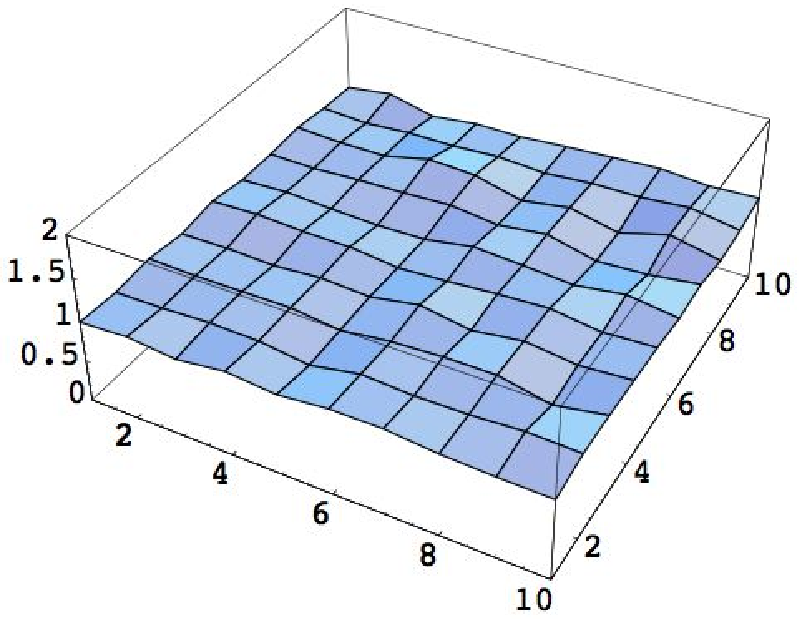}} \quad
	\subfigure[Sampled from $g(\mathbf{x},\theta)$]{\includegraphics[width=1.5in]{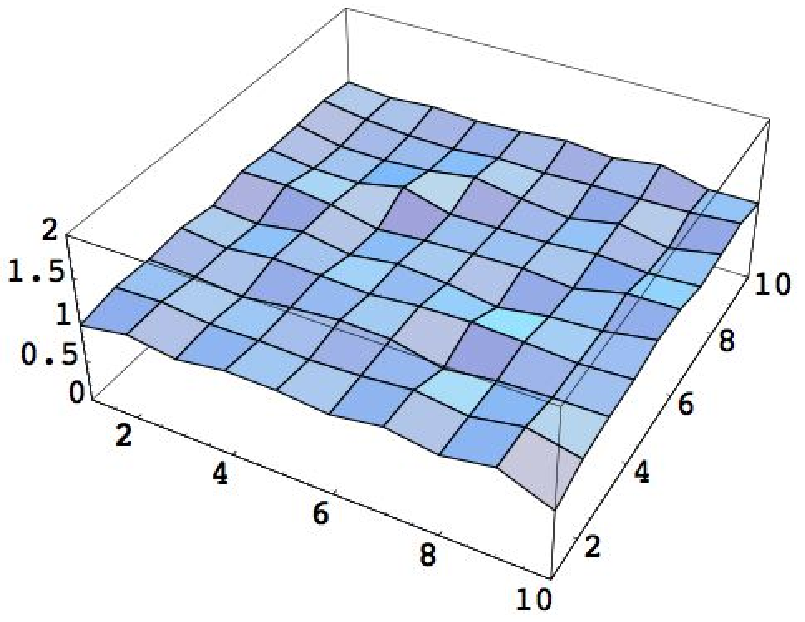}}
	}
   \end{center}
   \caption{Simulated mobile node spatial distribution function for $40,000$ nodes moving according to the RD model, with $v_{min}=0.001$m/s, $v_{max}=.01$m/s, $\tau_{pause}=1$s, $\tau_{move}=3$s, for $9,900$ s of simulated time on the unit ($1\times1$m$^{2}$) square, with different border rules.  Plotted is the node distribution at $t=9,900$ seconds.  mobile nodes that hit the border are: (a) re-introduced uniformly at random inside the square, (b) forced to choose a new direction that takes them back into the square, (c) wrapped around to the far side of the square, or (d) re-introduced to the square from a point on the border randomly sampled from the distribution (\ref{eq:finalpdf}).  Note that the x and y axes vary from $1$ to $10$, as we have actually plotted the contents of $100$ bins, each with area $.01$m$^{2}$. }
   \label{fig:RDSims}
\end{figure}

We show the mobile node distributions that result from each of the border rules discussed in this paper in Figure \ref{fig:RDSims}.  In Figure \ref{fig:RDSims}(a), we see that replacing mobile nodes uniformly at random within the square produces a markedly nonuniform distribution.  Figures \ref{fig:RDSims}(b) and (c) show the uniform distributions expected when nodes are (b) reflected or (c) wrapped at the border.  Figure \ref{fig:RDSims}(d) shows the uniform distribution that results from the application of our new border rule.

Additional realism can be obtained by relaxing the condition, required for some protocol simulation frameworks, that the number of nodes within the box be constant.  Instead, mobile nodes can be introduced at random intervals with a rate equal to the rate at which mobile nodes exit the test area.  This more accurately mimics the motion of mobile nodes in a larger test area moving without regard to the arbitrarily defined simulation geometry.  One possible complication is that this would cause the total number of mobile nodes in the test area to fluctuate with time (typically on the order of $\sqrt{N}$, $N$ the average number of mobile nodes simulated), and thus testbeds may have to be modified to take full advantage of this approach.

\section{Uniformity in the Random Waypoint Model\label{sec:rwp}}

The random waypoint (RWP) model is often used to model intentional movement.  One drawback of the RWP model is that the steady-state mobile node distribution is nonuniform.  There has been some interest in finding ways to increase the uniformity of the RWP mobile node distributions \cite{Bettstetter:2002,Blough:2004}.  Here, we examine what makes the RWP distributions nonuniform, and prove that the RWP model can never produce uniform mobile node distributions.  We also show that the RWP model is a limiting case of an RD model with a ``volume rule''.  In constructing this volume rule, we propose the use of a hybrid RWP model that can help (but never fully) flatten the steady-state mobile node distribution.

\subsection{How the RWP model leads to nonuniform distributions}
In \cite{Blough:2004}, it is observed that for simulations of the RWP model with a defined $p_{stat}$ (the probability that a node, initially uniformly distributed in the box, never moves), allowing for an average pause time $\tau_{pause}$ between movements, that only $p_{stat}$ and $\tau_{pause}$ influence the steady-state node distribution function.  

This result can be understood in the following way.  At any time, mobile nodes in the RWP model belong to one of three classes: nodes that never move, nodes that are pausing between steps, and nodes that are moving.  The initial node distribution is uniform, so mobile nodes in the first class are distributed uniformly.  Similarly, the points between which nodes move are selected uniformly at random, so mobile nodes that are pausing before moving towards their destination are also uniformly distributed.  The overall nonuniformity is due solely to the mobile nodes that are moving.  Increasing $p_{stat}$ and $\tau_{pause}$ does nothing but increase the proportion of mobile nodes that are stationary at any given time.  Indeed, for any choice of $p_{stat}$ and $\tau_{pause}$, a fraction $p_{stat}+\frac{\tau_{pause}}{\tau_{pause}+\tau_{move}}$ of the mobile nodes is stationary.  Increasing the proportion of stationary nodes flattens the mobile node distribution, but at the cost of impairing the node mobility.

The authors of \cite{Blough:2004} also observe that the nonuniformity of the steady-state mobile node spatial distribution function is independent of $v_{max}$, the maximum mobile node speed.  This can be understood via a one-dimensional model.  Consider a mobile node that moves according to the RWP model on a 1-D unit line segment between $0$ and $1$.  We know that the mobile node contributes to the nonuniformity of the distribution only if it is moving, so we examine the distribution to which $x_{obs}$, the observed position of a moving mobile node, belongs.  Ideally, we would like to find a way to make this distribution uniform.  Given that the node is traveling between points $x_{1}$ and $x_{2}$, the observed position of the mobile node is given by:
\begin{equation}
x_{obs}=x_{1}+(x_{2}-x_{1})f,
\end{equation}
where $f$ is the fraction of the mobile node's trajectory that has been completed.  If the mobile node moves at a constant speed from $x_{1}$ to $x_{2}$, and the time that we sample the mobile node's position is uncorrelated with the motion of the mobile node, $f$ is uniformly distributed on the interval $[0,1]$\footnote{See also \cite{Navidi:2004} where the uniformity of f is used to construct the stationary state.}.  If we average over initial and final points $x_{1}$ and $x_{2}$, we find
\begin{equation}
\langle x_{obs}\rangle = \frac{1}{2},
\end{equation}
which is consistent with $x_{obs}$ belonging to a uniform distribution.  That $x_{obs}$ has the same mean as a uniformly distributed variable on the interval $[0,1]$ does not however, prove that it is uniformly distributed.  In order for that to be so, \emph{all} moments of the distribution to which $x_{obs}$ belongs must be equal to the moments of a uniform distribution.  On $[0,1]$, the moments of a uniform distribution are $\langle x\rangle=1/2$, $\langle x^{2}\rangle=1/3$, $\langle x^{3}\rangle=1/4$, $\dots$, $\langle x^{n}\rangle=1/(n+1)$.  It behooves us to check the second moment of the distribution of $x_{obs}$, which is given by
\begin{equation}
\langle x_{obs}^{2}\rangle=\frac{1}{3}-\frac{1}{6}(f-f^{2}).\label{eq:xobsmoment2}
\end{equation}
The only way that the second moment of the distribution of $x_{obs}$ could be $1/3$ is if the second term in (\ref{eq:xobsmoment2}) were zero.  This could only happen if $f$ were infinitely sharply peaked at $f=0$ and at $f=1$.  Because there is a finite (typically large) probability that a moving node will be seen at points between the uniformly distributed endpoints, there is no way for $\langle x_{obs}^{2}\rangle$ to be uniformly distributed on $[0,1]$.  In particular, if $f$ is uniformly distributed on $[0,1]$, $\langle x_{obs}^{2}\rangle=\frac{11}{36}\simeq0.3056$.  Thus there is no way to fully eliminate nonuniformities in RWP models.

\subsection{Understanding and Improving the RWP model}
One can relate the RWP model to a more general mobility model defined on all space.  The RWP model has no explicit border rule, since mobile nodes are restricted to travel between points inside the test region.  Despite this, we can reproduce the RWP model via the application of a ``volume rule'' \footnote{The space in which mobile nodes move is an n-dimensional volume, where n is the number of spatial degrees of freedom.  In this context, we use ``volume'' to refer to the 2-dimensional space available to the mobile nodes.} to an RD model on a larger space.  This is simply the RD model of section \ref{sec:diffusion} defined on a larger area $\mathcal{V}$, a square with side $L$.  The actual test region is $\mathcal{A}\subset\mathcal{V}$, a square with side $d$, with $d\ll L$.  We choose $\tau_{m}\langle v\rangle\sim fL$, where $f\in(0,1]$, and $\langle v\rangle$ is the mea mobile node speed.  A border rule that produces uniform steady-state mobile node distributions is applied whenever mobile nodes leave $\mathcal{V}$.  To reproduce the RWP model, we define:

\begin{quote}
RWP Volume Rule

After selecting a set $\{\theta,t_{s},t_{m},v\}$ we (a) check to see if the destination point defined by $t_{m},v,\theta$ lies within $\mathcal{A}$.  (b) If so, the mobile node will move towards it as indicated by the RD model, and the rule has been applied.  (c) If not, the mobile node is instantly moved outside $\mathcal{A}$, and replaced with an identical mobile node that selects a new set $\{\theta,t_{s},t_{m},v\}$, and return to (a).  (c) ensures that the mobile node selects a destination that lies within $\mathcal{A}$.
\end{quote}

In the limit that $L\rightarrow\infty$, and in particular as $fL\gg d$, the probability that mobile nodes inside the test region will select a set $\{\theta,t_{s},t_{m},v\}$, which yields a destination point within the test region is roughly equal for all points within the test region.  At the same time, no mobile nodes will move across the boundary of the test region.  The RWP model has thus been reproduced via application of a volume rule to an underlying RD model.  We note, however, that this volume rule allows mobile nodes to move between points instantaneously, violating condition \ref{cond:two}.\footnote{Alternatively, the volume rule may be considered to bias the direction in which mobile nodes move, violating condition \ref{cond:three}}  Thus the distributions generated by the RWP model do not obey the diffusion equation.

This formulation of the RWP model draws attention to the source of the RWP model's problems.  mobile nodes that choose destinations outside the test region are removed instantaneously, and are not allowed to move continuously to the border.  At the same time, replacement mobile nodes do not move in from the border, but are simply dropped into the test region.  The exclusion of this kind of mobile node motion artificially depletes the areas near the border.  If we allow mobile nodes to enter and exit the test region according to the underlying RD model, this depletion can be eliminated.  Of course, including this motion while simultaneously setting $fL\gg d$ would cause mobile nodes to visit only a single point within the test region before leaving it.

This suggests a compromise.  The RWP model is considered to be realistic because mobile nodes move in straight lines to points throughout the test region, rather than in random walks that slowly move away from an initial point.  We can duplicate this desirable behavior by increasing $\tau_{m}$ in the RD model.  As $\tau_{m}\langle v\rangle$ increases, the mobile node is equally likely to move to any point within the test region.  It is desirable that our model cause mobile nodes to visit multiple points within the test region, so we randomly apply the RWP volume rule with probability $p_{confine}$ at the beginning of each step.  In the long run, the unconfined fraction will serve to reduce the RWP depletion of the borders.  We have performed such a simulation for $40,000$ mobile nodes with the parameters $v_{min}=0.1$m/s, $v_{max}=1.0$m/s, $\tau_{s}=0$s, $\tau_{m}=50,000$s, on a test region $\mathcal{A}$ defined by a $1\times1$m$^{2}$ square, using the new border rule introduced in section \ref{sec:distsec}.  We optimistically determine the mobile node boundary distribution by placing mobile nodes uniformly at random within the test region and running the simulation forward one step.  To save some time running the simulations, we allow the mobile nodes incident upon the border from the outside to choose an direction $\theta$ uniformly at random, so long as it enters the test region.  We vary the value of $p_{confine}$ from $1$ to $0.2$.  After $2,000$ seconds of simulated time, we see that as $p_{confine}$ decreases, the distribution of mobile nodes flattens out.  We also notice that when $p_{confine}=0.2$, the characteristic bump of the RWP model has become a depression.  This is a consequence of our choosing to assume the distribution of angles at which mobile nodes enter the test region is uniform, and is another example of the importance of choosing the correct boundary conditions in designing simulations.
\begin{figure}[h]
\begin{center}
\mbox{
\subfigure[$p_{confine}=1$]{\includegraphics[width=1.5in]{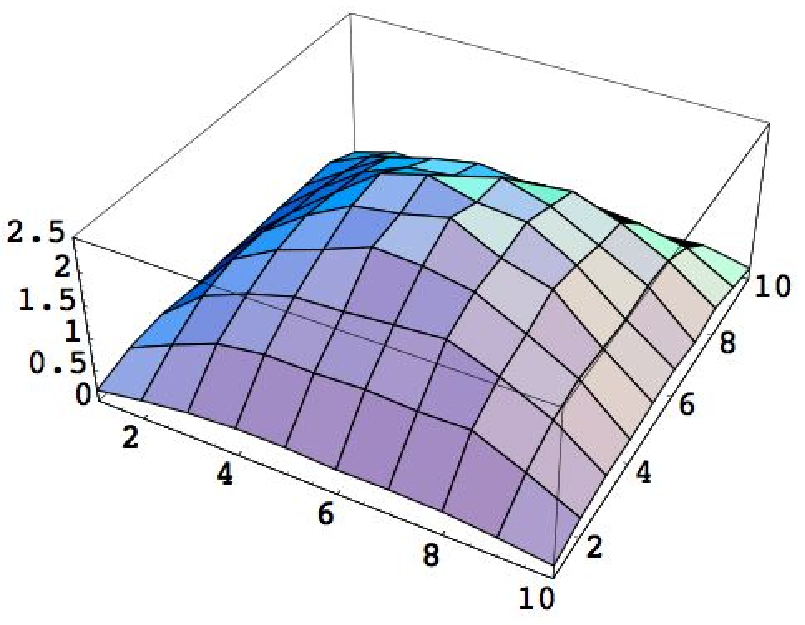}}\quad
\subfigure[$p_{confine}=0.8$]{\includegraphics[width=1.5in]{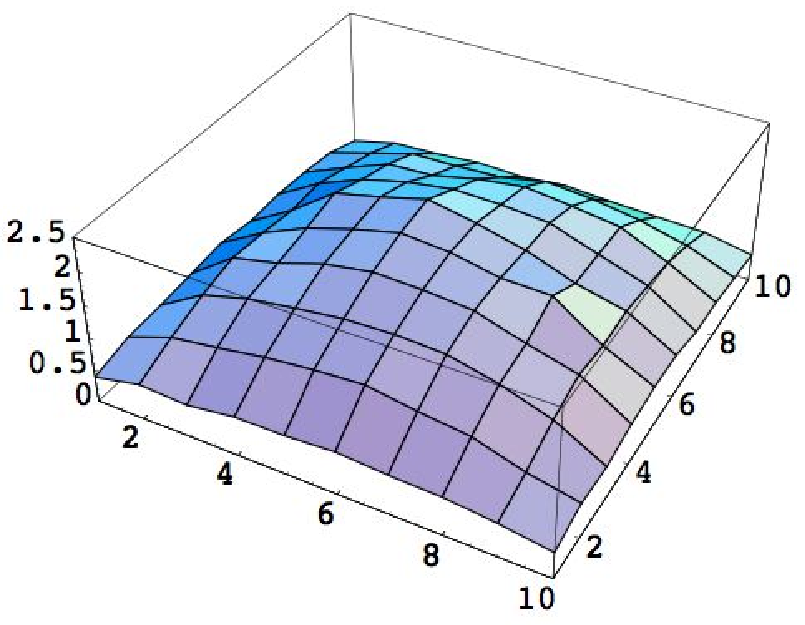}}
}
\mbox{
\subfigure[$p_{confine}=0.6$]{\includegraphics[width=1.5in]{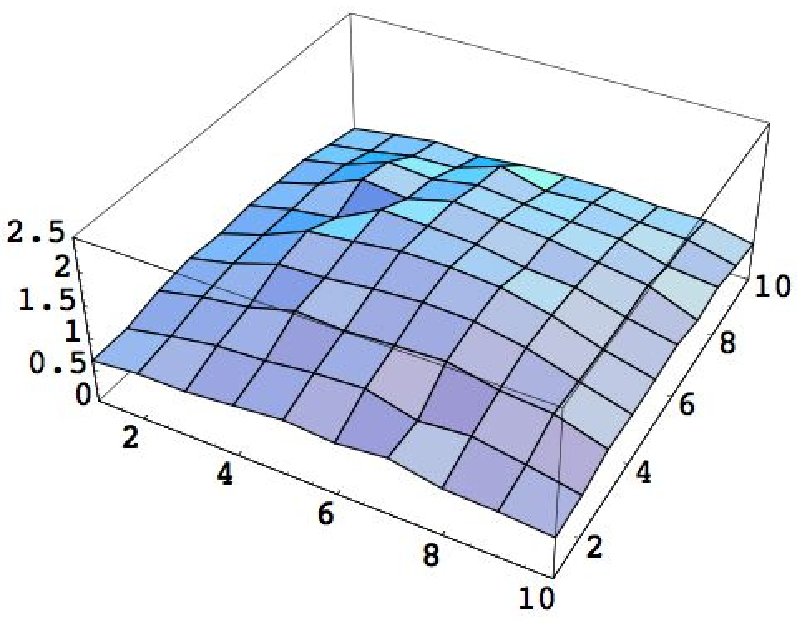}}\quad
\subfigure[$p_{confine}=0.4$]{\includegraphics[width=1.5in]{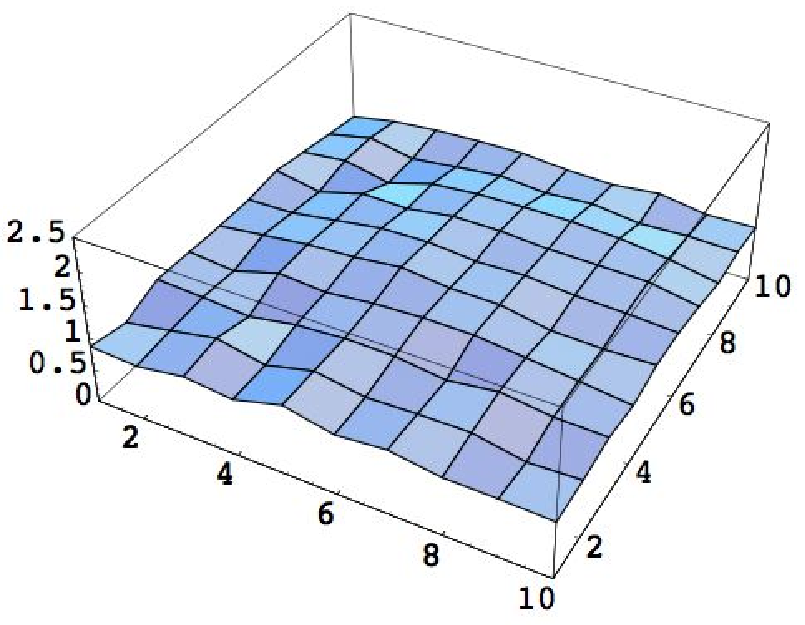}}\quad
\subfigure[$p_{confine}=0.2$]{\includegraphics[width=1.5in]{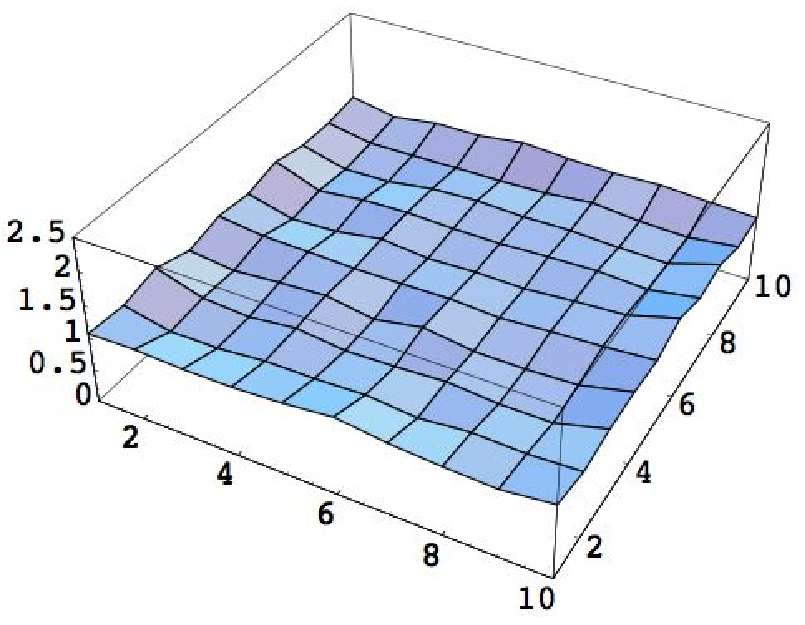}}
}
\end{center}
\caption{Simulated distribution of mobile nodes for $40,000$ nodes moving according to the hybrid RWP model proposed in section \ref{sec:rwp} for $2,000$ seconds.  $v_{min}=0.1$m/s, $v_{max}=1.0$m/s, $\tau_{s}=0$s, $\tau_{m}=50,000$s.  Shown are the distributions that result for different values of $p_{confine}$.  $p_{confine}=$ (a) $1.0$, (b) $0.8$, (c) $0.6$, (d) $0.4$, and (e) $0.2$.}
\label{fig:RWPDists}
\end{figure}

\section{Conclusion}

We have demonstrated a new border rule for use in simulating node mobility.  This rule produces mobile node distributions that do not depend upon the particulars of our simulated space.  It improves upon prior techniques chiefly in that it does not rely upon unrealistic reflection or wrapping of mobile nodes at the boundary of the test region, and mimics the way mobile nodes would arrive and depart the test region if they are assumed to obey a given mobility model in all space.  In particular, we have demonstrated this rule for the random direction (RD) model.

We have shown that most mobility models in current use produce mobile node distributions that obey the diffusion equation.  The random waypoint (RWP) model is an exception to this, but we have shown that the RWP model can be understood as a limiting case of an RD model, where a ``volume rule'' has been applied.  This volume rule is responsible for violating the conditions under which mobile nodes obey the diffusion equation.

We have also shown that the RWP model can never yield uniform mobile node distributions.  We have proposed a ``hybrid'' RWP model, which combines the desirable ``intentional'' motion of the RWP model with a way to increase the node density near the borders, allowing us to partially mitigate the border effect.

\section{Acknowledgements}
The author would like to thank Itay Yavin and Jennifer Ellsworth for many fruitful discussions related to this work, as well as Paul Hohensee and Victor Luchangco for helpful comments on this work.

\appendix
\section{The marginal p.d.f.\label{sec:analyticDist}}
We now demonstrate how to determine the marginal probability density function $g(\mathbf{x})$.

For an arbitrary point on the border, defined by the vector $\mathbf{x}$, the distance between this point and any visible point $\mathbf{x}'$ \footnote{a point $\mathbf{x}'$ is ``visible'' to $\mathbf{x}$ if the line of travel between the two points does not contain points outside the test region.} on the interior of the test region is $d(\mathbf{x},\mathbf{x}')$.  We assume the distribution of nodes to be uniform at $t=0$ and for all time thereafter.  We want to know the rate at which mobile nodes impinge upon the point $\mathbf{x}$ as a function of the angle of incidence with the surface, and the distance traveled to reach $\mathbf{x}$.  This rate is equal to the rate at which a resting mobile node will start to move, $1/\tau_{s}$, multiplied by the probability that a mobile node will cover the distance $d(\mathbf{x},\mathbf{x}')$ between the border and its starting point in a single move.  That is:
\begin{equation}
\frac{P(\text{to cover } d(\mathbf{x},\mathbf{x}'))}{\tau_{s}}. \label{eq:rateimpinge}
\end{equation}
Thus the total rate at which mobile nodes leave the test region through $\mathbf{x}$ at a particular angle $\theta$ is simply the line integral of (\ref{eq:rateimpinge}) from $\mathbf{x}$ to the furthest point $\mathbf{x}'$ in the test region along this direction (i.e. a point on the far border).  That is to say, the rate $R$ is
\begin{align}
R(z(\mathbf{x},\theta))&=\int_{0}^{z(\mathbf{x},\theta)}\frac{P(\text{to cover } d(\mathbf{x},\mathbf{x}'))}{\tau_{s}}d\mathbf{l} & z(\mathbf{x},\theta)&=\max d(\mathbf{x},\mathbf{x}'),\label{eq:incidentflux}
\end{align}
where we have explicitly noted that the furthest distance a mobile node could have traveled in a single movement is a function of the direction we consider, and is equal to the distance from $\mathbf{x}$ to another visible point on the border that lies along the direction $\theta$.  We assume the border to be piecewise smooth and continuous, and with no loss of generality, we will restrict the analysis presented here to a two-dimensional test region.  We define $\theta$ to be the angle that a given mobile node trajectory makes with the normal to the border at $\mathbf{x}$.  For smooth regions, $\theta\in[-\pi/2,\pi/2]$.  For points in the set of corners, $\theta$ may vary over a different domain, but because the border is assumed to be piecewise smooth, this set has zero measure, and thus the probability that a mobile node will exit through one of these points is negligible.  Our final specification is that our test region be a unit square.  In this case, symmetry reduces the problem to finding the rate at which mobile nodes leave the test region through a single side of the box.  

The distance to the far side of the box as a function of $x$, the position of the chosen exit point on the side of the box, and the angle $\theta$ with the normal to the surface of the box, is
\begin{equation}
z(x,\theta) = \begin{cases}
\frac{x}{\sin\theta} & \text{for $\theta\in\left[-\pi/2,-\tan^{-1} x\right]$}\\
\frac{1}{\cos\theta} & \text{for $\theta\in\left[-\tan^{-1}x,\tan^{-1}(1-x)\right]$}\\
\frac{1-x}{\sin\theta} & \text{for $\theta\in\left[\tan^{-1}(1-x),\pi/2\right]$}
\end{cases}
\end{equation}

If we know the probability that a mobile node will travel a distance $d$ before stopping, we can find the rate at which they leave through a small region about $x$.

For the RD mobility model, the distance a mobile node moves is determined by the time it moves before stopping (assumed to be exponential with mean $\tau_{m}$) and the velocity at which it moves (assumed to be uniformly distributed between $v_{min}$ and $v_{max}$).  Thus we begin with the p.d.f. over time and velocity:
\begin{equation}
f(t,v) = \frac{e^{-t/\tau_{m}}}{\tau_{m}(v_{max}-v_{min})}
\end{equation}
We wish to express this p.d.f. in terms of $d$ and $v$, so we perform the change of variables $d = t v$, $v=v$, with Jacobian $\left[\begin{array}{cc}v & t\\ 0&1\end{array}\right]$.
\begin{equation}
f(d,v) = \frac{e^{-d/\tau_{m}v}}{v\tau_{m}(v_{max}-v_{min})}
\end{equation}
The probability that a mobile node will cover a distance of at least $D$ is thus:
\begin{multline}
P(\text{to cover } D)=\int_{D}^{\infty}dd\int_{v_{min}}^{v_{max}}dv \:f(d,v)=\\
\frac{1}{\tau_{m}(v_{max}-v_{min})}\left[\tau_{m}\left(v_{max}e^{-\frac{D}{v_{max}\tau_{m}}}-v_{min}e^{-\frac{D}{v_{min}\tau_{m}}}\right)-D\left[\Gamma\left(0,\frac{D}{v_{max}\tau_{m}}\right)-\Gamma\left(0,\frac{D}{v_{min}\tau_{m}}\right)\right]\right]
\end{multline}
and the rate at which mobile nodes reach the point $\mathbf{x}$ at an angle $\theta$ is from (\ref{eq:incidentflux}),
\begin{multline}
R(z(\mathbf{x},\theta))=\frac{\tau_{m}}{2\tau_{m}\tau_{s}(v_{max}-v_{min})}\Bigg[(v_{max}^{2}-v_{min}^{2})\tau_{m}+v_{max}(z(\mathbf{x},\theta)-v_{max}\tau_{m})e^{-\frac{z(\mathbf{x},\theta)}{v_{max}\tau_{m}}}\\
-v_{min}(z(\mathbf{x},\theta)-v_{min}\tau_{m})e^{-\frac{z(\mathbf{x},\theta)}{v_{min}\tau_{m}}}-(z(\mathbf{x},\theta)^{2}/\tau_{m})\left[\Gamma\left(0,\frac{z(\mathbf{x},\theta)}{v_{max}\tau_{m}}\right)-\Gamma\left(0,\frac{z(\mathbf{x},\theta)}{v_{min}\tau_{m}}\right)\right]\Bigg].
\end{multline}
Alternatively, if we define $d_{l}=v_{min}\tau_{m}$ and $d_{h}=v_{max}\tau_{m}$, and explicitly express $R$ as a function of $\mathbf{x}$ and $\theta$:
\begin{multline}
R(\mathbf{x},\theta)=\\
\frac{(d_{h}^{2}-d_{l}^{2})+d_{h}(z(\mathbf{x},\theta)-d_{h})e^{-\frac{z(\mathbf{x},\theta)}{d_{h}}}-d_{l}(z(\mathbf{x},\theta)-d_{l})e^{-\frac{z(\mathbf{x},\theta)}{d_{l}}}-z(\mathbf{x},\theta)^{2}\left[\Gamma\left(0,\frac{z(\mathbf{x},\theta)}{d_{h}}\right)-\Gamma\left(0,\frac{z(\mathbf{x},\theta)}{d_{l}}\right)\right]}{2\tau_{s}(d_{h}-d_{l})}.
\end{multline}
Given the specific geometry of the boundary, we may write $z(\mathbf{x},\theta)$, and we may then integrate $R(z(\mathbf{x},\theta))$ over $\theta$ to obtain the unnormalized p.d.f. for the point at which exiting mobile nodes leave the box:
\begin{equation}
f(x)=\int_{-\pi/2}^{-\tan^{-1}x}R\left(\frac{-x}{\sin\theta}\right)d\theta+\int_{-\tan^{-1}x}^{\tan^{-1}(1-x)}R\left(\frac{1}{\cos\theta}\right)d\theta+\int_{\tan^{-1}(1-x)}^{\pi/2}R\left(\frac{1-x}{\sin\theta}\right)d\theta.\label{eq:finalpdf}
\end{equation}

In this case, there is no closed form solution to eq. (\ref{eq:finalpdf}).  Numerical evaluation yields the p.d.f. whose cumulative distribution function is shown by the solid line in Figure \ref{fig:RDDists}.

\bibliography{BoundaryEffectPaper}
\bibliographystyle{unsrt}
\end{document}